# Development of a photonic crystal spectrometer for greenhouse gas measurements

Marijn Siemons[a], Martijn Veen[a], Irina Malysheva [a,b], Johannes Algera[a,c], Stefan Philippi[a,b], Kirill Antonov[d], Niki van Stein[d], Jérôme Loicq[c], Nandini Bhattacharya[b], René Berlich[e], Anna V. Kononova[d], Ralf Kohlhaas*[a,b]

[a]SRON Netherlands Institute for Space Research, 2333 CA Leiden, The Netherlands; [b]Department of Precision and Microsystems Engineering, Delft University of Technology, 2628 CD Delft, The Netherlands; [c]Faculty of Aerospace Engineering, Delft University of Technology, 2629 HS Delft, The Netherlands; [d]Leiden Institute of Advanced Computer Science, Leiden University, 2333 CA Leiden, The Netherlands; [e]European Space Agency, ESTEC, 2200 AG Noordwijk, The Netherlands

**ABSTRACT**

The need of atmospheric information with a higher spatial and temporal resolution drives the development of small satellites and satellite constellations to complement satellite flagship missions. Since optical systems are a main contributor to the satellite size, these are the prime candidate for their miniaturization. We present here a novel optical system where the complete spectrometer part of the optical system is compressed in one flat optical element. The element consists of an array of photonic crystals which is directly placed on a detector. The photonic crystals act as optical filters with a tunable spectral transmission response. From the integrated optical signals per filter and the atmosphere model, greenhouse gas concentrations are obtained using computational inversion. We present in this article the instrument concept, the manufacturing and measurement of the photonic crystals, methods for the filter array optimization, and discuss the predicted retrieval performance for the detection of methane and carbon dioxide.

## 1. INTRODUCTION

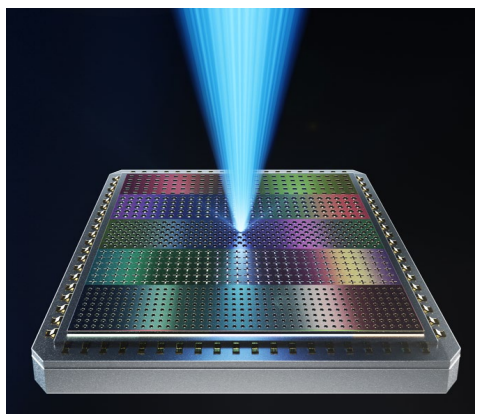

Figure 1: Artist impression of photonic crystal array integrated on a detector

Current satellite flagship trace gas measurement missions such as the Sentinel-5 precursor TROPOMI [1] typically rely on dispersive optical elements to spatially separate wavelengths on a detector. In case of grating spectrometers, this leads to complex optical systems with a high physical volume due to the need of the optical propagation space. In contrast, systems based on interference filters [2,3] or Fabry-Perot filters [4] can be more compact and can operate with a larger field-of-view in the along-flight direction. However, inherently to bandpass filter systems a large fraction of the incident light is reflected. This means that effectively not all light collected by the telescope is used for trace gas retrieval. Concepts based on static Fourier transform spectrometers [5] can have a higher throughput, but only have a limited number of available transmission functions.

*r.kohlhaas@sron.nl

Inspired by the recent invention of on-chip photonic crystal spectrometers [6], we explore an instrument concept where photonic crystals are placed on a detector and used as complex transmission filters with a high total transmission. With the known filter transmission profiles and a-priori knowledge of the atmosphere model a trace gas abundancy is determined. We give in this proceeding article an overview of the current development status, explaining the instrument concept, report on photonic crystal manufacturing and testing, filter selection, preliminary telescope design and give a first performance estimate for a system for methane and carbon dioxide detection.

## 2. INSTRUMENT CONCEPT

An overview of the instrument concept can be found in Figure 2. The optical system (A) consists of a telescope which images a ground scene on a detector with photonic crystals. The satellite flies in low-earth orbit in push-broom configuration (B). The photonic crystal filters are oriented in stripes (C) perpendicular to the across-track (ACT) direction. In this way during the along-track (ALT) flight of the satellite the same ground elements are seen by different filter functions. The photonic crystals (D) consist of periodic hole patterns such as circles, squares or crosses in a dielectric material. We focus at first on the Short-Wave Infrared (SWIR) spectral range from 1.6 µm to 1.7 µm for the detection of methane and carbon monoxide. In this spectral range silicon is an attractive photonic crystal material due to its high refractive index and low absorption. We place the photonic crystals on a glass substrate. In the regime where wavelength of the light is of the order of the photonic crystal period, there is a strongly wavelength dependent transmission spectrum (E). By tuning the photonic crystal parameters, different transmission functions can be selected.

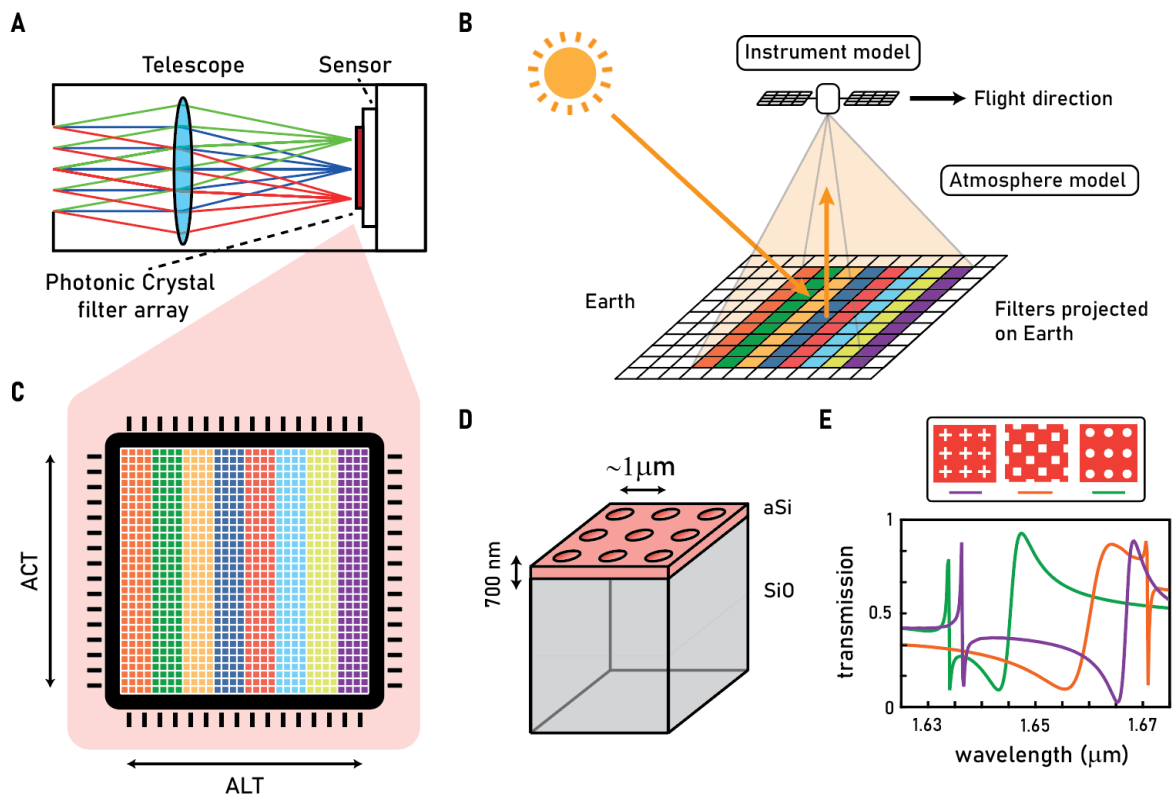

Figure 2: Overview of instrument concept. (A) A telescope images a ground scene on a detector. (B) The satellite flies in push-broom configuration over the earth. (C) Photonic crystal filters arranged as stripes on the detector. (D) Illustration of photonic crystal structure. (E) Different photonic crystal transmission spectra for different photonic crystals.

Due to the higher ALT field-of-view and the possibility to choose the measurement basis with the photonic crystal transmission profiles, a higher performance in comparison with a classical imaging grating spectrometer can be reached. The trace gas retrieval relies on the direct integration of the instrument and retrieval model. The signal per filter k for one ground element is given by

$$\mu_k = t_{\text{int}} G N_{\text{pxl}} \int_{\lambda_{\min}}^{\lambda_{\max}} QE(\lambda) T_k(\lambda) S(\lambda) d\lambda$$

where is $t_{\text{int}}$ is the integration time per image, G the etendue, QE the quantum efficiency of the detector, $T_k$ the filter transmission function, $N_{\text{pxl}}$ the number of pixels used, S the spectral radiance of the earth and $\lambda$ the wavelength. For a first instrument design, we aim at a spatial resolution of 300 m × 300 m at an altitude of 500 km, the same as the current baseline of the TANGO Scout mission [7]. As a detector a Lynred Snake InGaAs detector is chosen. The detector has 512 × 640 pixels, a full well capacity of $1.4 \cdot 10^6$ e$^-$, a pixel size of 15 µm and an estimated QE of 85%. The focal length (125 mm) is chosen for a ground instantaneous field-of-view of 60 × 60 m². An integration time of 34 ms is set for a smear for an along-track pixel size of 300 m. 5 ACT pixels are binned for a 300 × 300 m² resolution element and a total across track swath of 30 km. This design results in a F-number of 6.1 and an etendue of $5.8 \cdot 10^{-12}$ m²sr.

The earth's atmosphere is modelled with a non-scattering radiative transfer model. The spectral radiance is given by

$$S(\lambda) = \frac{\cos(\zeta_s)}{\pi} E(\lambda) A(\lambda) \exp\left[-r_{\text{air}} \tau_{\text{vert}}(\lambda)\right]$$

Where $\zeta_s$ is the sun zenith angle, $E$ the sun's irradiance, $A$ the albedo of the earth, $r_{\text{air}}$ the air mass factor and $\tau_{\text{vert}}$ the vertical optical thickness. The spectral dependance of the albedo of the earth is expressed a polynomial and we use here 2 albedo terms. The trace gas abundancies and albedo terms are determined with the known filter functions with a least-squares fit

$$\min_\theta \left(n_k - \mu_k(\theta)\right)^2$$

where θ are the fit parameters (atmospheric gases, albedo terms), $n_k$ is the measured photon count for filter k, and $\mu_k$ the expected model photon count. With this model the trace gas abundancies can be retrieved. For an accurate measurement of emission plumes with a non-scattering radiative transfer model, the proxy method [8] is used which assumes non-mixed sources. For mixed sources an additional instrument for light path correction such as a multi-angle polarimeter would be needed [9].

## 3. PHOTONIC CRYSTAL MANUFACTURING

The photonic crystals are manufactured by electron beam (e-beam) writing and Reactive Ion Etching (RIE). Amorphous silicon with a thickness of 400 nm is deposited on 500 µm thick fused silica wafer with Inductively Coupled Plasma

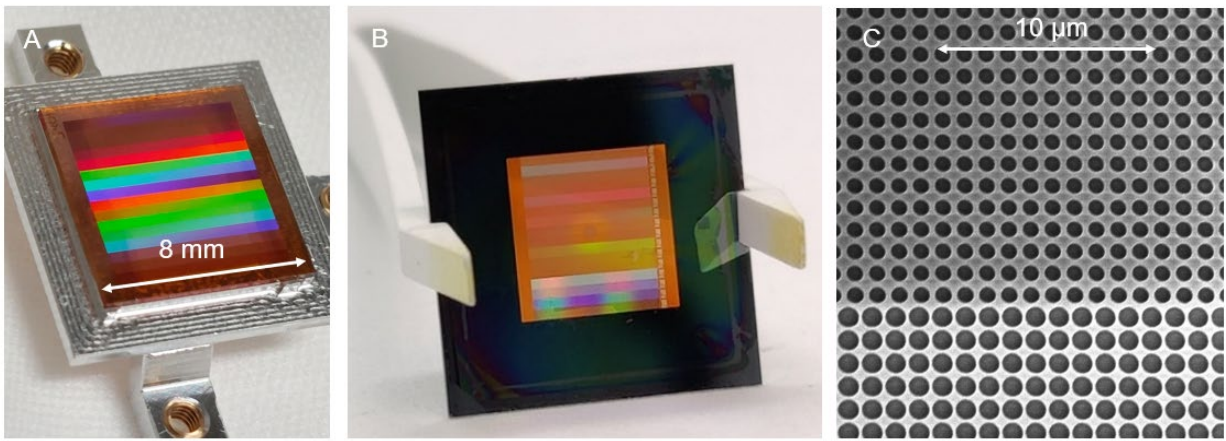

Figure 3: Manufactured photonic crystal arrays. (A) Photonic crystal array on glass. (B) Photonic crystal array on a SiN membrane. (C) Zoom-in with a scanning electron microscope on the sample from (B).

Chemical Vapour Deposition (ICP-PECVD). A positive e-beam resist (CSAR) is spinned on the surface and covered with a thin chromium layer to reduce charging effects. After e-beam writing, the chromium layer is stripped, and the resist is developed. RIE is done with an SF6-O2 plasma. Finally, the resist is stripped, and the samples are diced matching to the detector size. After dicing, the photonic crystal arrays are glued on a mechanical adaptor. In an additional alternative manufacturing method, silicon photonic crystals are manufactured on a silicon nitride membrane. Instead of on glass, amorphous silicon is deposited on a silicon wafer with a silicon nitride layer. After e-beam writing and development a large hole is etched on the backside of the silicon wafer to obtain a silicon nitride membrane of 1000 nm thickness. In Figure 3 A, a full photonic crystal array on glass, and in Figure 3 B a silicon photonic crystal on a membrane can be seen. In Figure 3 C a zoom-in with a scanning electron beam microscope is shown.

## 4. OPTICAL TESTING

The photonic crystals are interferometrically aligned to the detector with a collimated beam. In this way a distance of a few tens of μm between the detector and the silicon photonic crystals surface is reached. The photonic crystal stripes are further aligned with the detector pixels. After alignment, the photonic crystals are measured with an image space telecentric lens system, see Figure 4. The output of a tunable laser passes a motorized despeckler and is fed into an integrating sphere. The output of the integrating sphere is imaged on the photonic crystal-detector system with an image space telecentric lens system. Changing the entrance aperture of the optical system allows to change the F-number of the system.

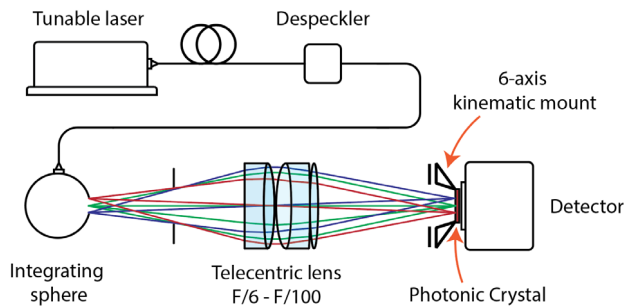

Figure 4: Image space telecentric optical system for the measurement of the photonic crystals.

A measurement of a photonic crystal array with patches of different photonic crystals with an area of 150 × 150 μm² is shown in Figure 5 A. The organization in patches allows the measurement of many different photonic crystals at once. In Figure 5 B an example measurement result with an F-number of F/25 is shown. Due to reflections between the air-glass surface and the photonic crystal surface Fabry-Perot fringes are formed. The fringes largely disappear when the F-number is decreased to F/6 and can be further reduced with an anti-reflection coating on the glass-air surface. In Figure 5 C the measurement with a collimated beam of a photonic crystal on a silicon nitride membrane is shown.

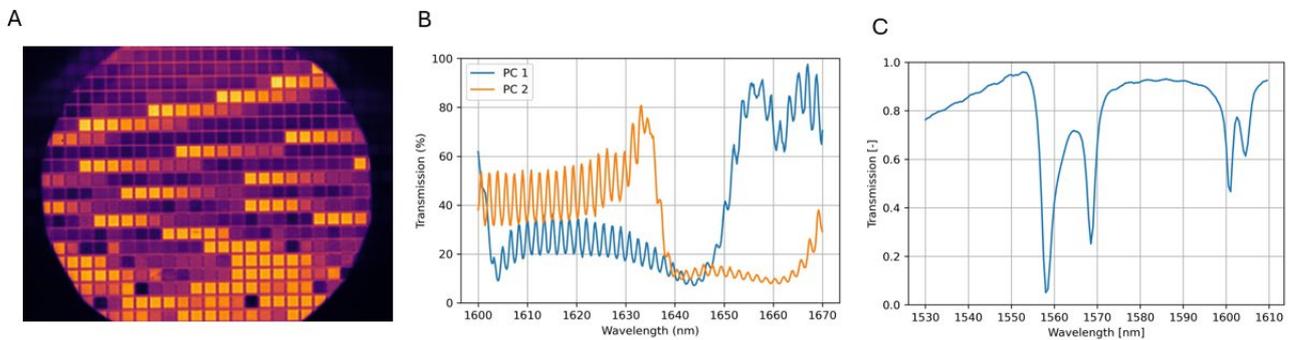

Figure 5: Transmission measurement of photonic crystal filters. (A) A 2D array with different photonic crystal patches. (B) Measurement of a photonic crystals on a glass substrate with a F/25 beam. (C) Measurement of a photonic crystal on a SiN membrane with a collimated beam.

## 5. FILTER SELECTION

For a computational photonic crystal spectrometer based on compressive sensing such as in [6], a filter set with a maximum spectrally uncorellated set of filters would be optimal. In contrast, we are using a method where the photonic crystal filter functions are directly integrated in the atmosphere retrieval model for trace gas retrieval. There is therefore a considerable a-priori knowledge which can influence the optimal filter choice. A large library of photonic crystal filters was simulated with a Rigorous Coupled-Wave Analysis RCWA software (Lumerical) and different filter selection methods were tested. Different tested filter selection methods included a maximum uncorrelated set, ranking by the second moment of the Fourier spectrum of the transmission functions, direct performance evaluation with help of the Cramer-Rao-Lower-Bound (CRLB) [10] and optimization with evolutionary algorithms. We preliminary choose a filter set containing 16 different filters. This is sufficient to cover the main modes in the atmospheric spectrum corresponding to the different gases in the spectral range and albedo terms while still offering redundancy. The CRLB-based method and evolutionary algorithms showed the best performance, while the CRLB method can be integrated into the evolutionary algorithms to improve the performance further.

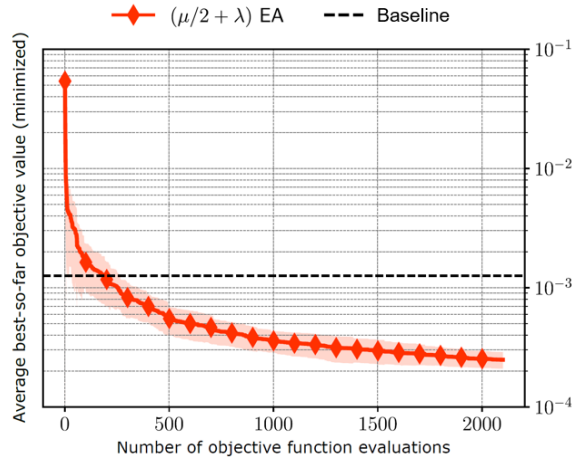

Figure 6: Filter set optimization with evolutionary algorithms. The optical performance with evolutionary algorithms (red) is compared to a baseline filter set.

Evolutionary algorithms (EA) [11] are nature-inspired algorithms that allow finding solutions of high quality but are not necessarily strictly optimal, within a limited computation budget. Such algorithms typically start with random solutions and iteratively look for improvements of such solutions. Every solution has the quality assigned by a fitness function, which defines the considered problem. Following the evolutionary paradigm, a population of solutions is evolved over a number of iterations. Typically, an iteration of EA consists of a number of steps: reproduction selection from the solutions in the population, producing new solutions with recombination of the selected solutions, mutation of the produced solution and survival selection to create a population in the subsequent iteration. The iterations are sequentially repeated until the termination criteria are met. In Figure 6, the filter set optimization with an evolutionary algorithm is shown. Each iterations contains steps such as mutation, crossover and selection. Evolutionary algorithms are stochastic and they are therefore executed a number of times as plotted as a shaded area in Figure 6. The red line is the average and shows the convergence of the optimization after 2000 iterations. The obtained filter set outperforms a baseline filter set obtained by the ranking of all filters in the library over the 2$^{nd}$ moment of the Fourier transform of the filter transmission spectra.

## 6. TELESCOPE DESIGN

In Figure 7, a preliminary design of a three-mirror anastigmat (TMA) corresponding to the instrument parameters from section 2 is shown. The design is image space telecentric and the total optical system volume is below one liter. This is approximately one order of magnitude lower than the volume of a comparable grating imaging spectrometer for a similar spatial resolution and swath. It indicates the potential of the concept for the volume reduction in comparison with state-of-the-art instruments.

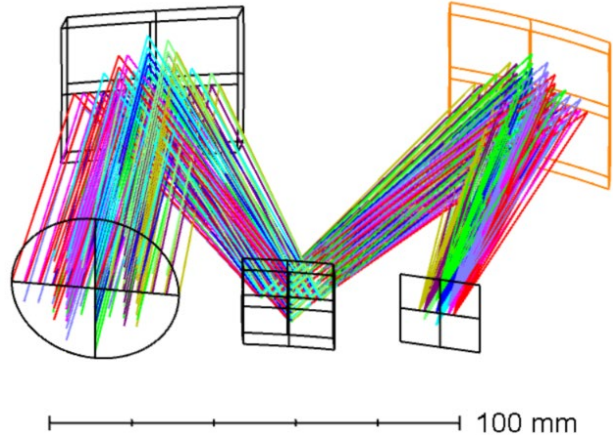

Figure 7: A preliminary three-mirror anastigmat (TMA) telescope design.

## 7. RETRIEVAL PERFORMANCE

The retrieval performance for an optimized filter set and a solar zenith angle (SZA) of 45 degrees shown in Figure 8. The retrieval error contains both precision and bias and it is before application of the proxy method. No forward motion compensation was assumed. Depending on the surface albedo, for methane a retrieval error between 0.4% to 1% and for carbon dioxide between 0.2% to 0.4% can be reached.

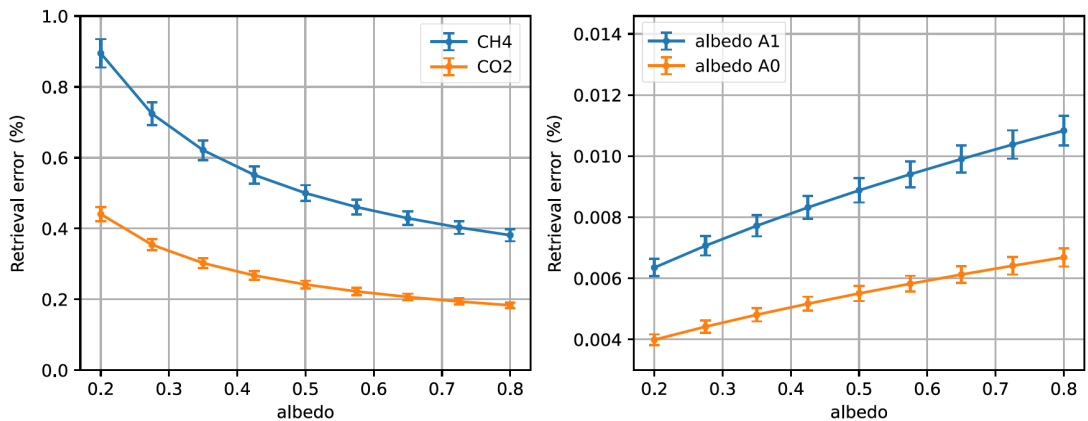

Figure 8: Retrieval performance for methane and carbon dioxide detection and surface albedo.

## 8. CONCLUSION AND OUTLOOK

In this article, we have given an overview of a photonic crystal instrument concept for earth observation. Greenhouse gases are retrieved with the help of photonic crystal filters. We showed that a similar performance as with a grating imaging spectrometer but with a significantly smaller instrument volume can be reached. In the next steps, we intend to scale the instrument concept to smaller spatial resolutions and aim for the full integration, optical testing, and environmental testing of a photonic crystal-detector module. We further envisage the application of the instrument concept for the detection of atmospheric gases at longer wavelengths, where the absolute gain in instrument volume in comparison with existing instrumentation could be even greater.

## ACKNOWLEDGEMENTS

We would like to thank ESA for financial support for this project with an Open Space and Innovation Platform system study. We thank Jochen Landgraf and Paul Tol for support on the retrieval code and Brecht Simon for support on the nanofabrication.

## REFERENCES

[1] de Vries, Johan, et al. "TROPOMI on ESA's Sentinel 5p ready for launch and use." Fourth international conference on remote sensing and geoinformation of the environment (RSCy2016). Vol. 9688. SPIE, 2016.

[2] Mattos, Bruno, et al. "Spectrometer for Methane Monitoring Using a Tilted Interference Filter." IGARSS 2023-2023 IEEE International Geoscience and Remote Sensing Symposium. IEEE, 2023.

[3] Blommaert, Joris, et al. "CSIMBA: Towards a smart-spectral cubesat constellation." IGARSS 2019-2019 IEEE International Geoscience and Remote Sensing Symposium. IEEE, 2019.

[4] Jervis, Dylan, et al. "The GHGSat-D imaging spectrometer." Atmospheric Measurement Techniques 14.3 (2021): 2127-2140

[5] Gousset, S., et al. "NANOCARB-21: a miniature Fourier-transform spectro-imaging concept for a daily monitoring of greenhouse gas concentration on the Earth surface." International Conference on Space Optics—ICSO 2016. Vol. 10562. SPIE, 2017.

[6] Wang, Zhu, et al. "Single-shot on-chip spectral sensors based on photonic crystal slabs." Nature communications 10.1 (2019): 1020.

[7] Day, James, et al. "Development of the TANGO carbon instrument for greenhouse gas detection." Remote Sensing Technologies and Applications in Urban Environments VIII. Vol. 12735. SPIE, 2023.

[8] Schepers, D., et al. "Methane retrievals from Greenhouse Gases Observing Satellite (GOSAT) shortwave infrared measurements: Performance comparison of proxy and physics retrieval algorithms." Journal of Geophysical Research: Atmospheres 117.D10 (2012).

[9] Rusli, Stephanie P., et al. "Anthropogenic CO 2 monitoring satellite mission: the need for multi-angle polarimetric observations." Atmospheric Measurement Techniques 14.2 (2021): 1167-1190.

[10] Huang, Fang, et al. "Video-rate nanoscopy using sCMOS camera–specific single-molecule localization algorithms." Nature methods 10.7 (2013): 653-658.

[11] Bäck, Thomas, Joost N. Kok, and G. Rozenberg. Handbook of natural computing. Springer, Heidelberg, 2012.